\documentclass{ws-jbcb}

\usepackage[utf8]{inputenc}
\usepackage{algorithm}     
\usepackage{algorithmic}     
\usepackage{hyperref}

\begin{document}

\markboth{Authors' Names}
{Cell Cycle and Protein Complex Dynamics in Discovering Signaling Pathways}

%
%

\title{Cell Cycle and Protein Complex Dynamics in Discovering Signaling Pathways}
%
%
\author{Daniel Inostroza}

\address{Computer Science Department, University of Concepci\'on,
\\Edmundo Larenas, Concepci\'on, 4030000, Chile\\
danielinostroza@udec.cl}
%
\author{Cecilia Hern\'andez\footnote{corresponding author}}

\address{Computer Science Department, University of Concepci\'on,
\\Edmundo Larenas, Concepci\'on, 4030000, Chile\\
Center for Biotechmology and Bioengineering (CeBiB), Santiago, Chile\\
cecihernandez@udec.cl}

\author{Diego Seco}

\address{Computer Science Department, University of Concepci\'on,
\\Edmundo Larenas, Concepci\'on, 4030000, Chile\\
dseco@udec.cl}

\author{Gonzalo Navarro}

\address{Center for Biotechnology and Bioengineering(CeBiB), \\ Santiago, Chile\\gnavarro@dcc.uchile.cl}

\author{Alvaro Olivera-Nappa}

\address{Center for Biotechnology and Bioengineering(CeBiB), \\ Department of Chemical Engineering and Biotechnology\\ University of Chile, Santiago, Chile\\
aolivera@ing.uchile.cl}

\maketitle

\begin{history}
\received{(Day Month Year)}
\revised{(Day Month Year)}
\accepted{(Day Month Year)}
\end{history}

\begin{abstract}
Signaling pathways are responsible for the regulation of cell processes, such as monitoring the external environment, transmitting information across membranes, and making cell fate decisions. Given the increasing amount of biological data available and the recent discoveries showing that many diseases are related to the disruption of cellular signal transduction cascades, in silico discovery of signaling pathways in cell biology has become an active research topic in past years. However, reconstruction of signaling pathways remains a challenge mainly because of the need for systematic approaches for predicting causal relationships, like edge direction and activation/inhibition among interacting proteins in the signal flow. We propose an approach for predicting signaling pathways that integrates protein interactions, gene expression, phenotypes, and protein complex information. Our method first finds candidate pathways using a directed-edge-based algorithm and then defines a graph model to include causal activation relationships among proteins, in candidate pathways using cell cycle gene expression and phenotypes to infer consistent pathways in yeast. Then, we incorporate protein complex coverage information for deciding on the final predicted signaling pathways. 
We show that our approach improves the predictive results of the state of the art using different ranking metrics.

\end{abstract}


\section{Introduction}
{P}{roteins} are molecules formed by sequences of amino acids. They usually interact with each other to perform specific functions in organisms. Discovering the protein roles in different functions is an important research area in the biological and biochemical fields, because of the impact that such information may have in the creation of new treatments of several diseases and in the comprehension of functions in living systems\cite{FunDis}. A kind of cell activity where several proteins work together in a temporal and spatial sequence is called ``signaling pathway''. A signaling pathway can be seen as a linear path in cascade, where multiple proteins associate or modify each other to perform a specific function. In general, a signaling pathway has a set of proteins whose sequence interaction from a source to a target produces the activation of transcription factors, which regulate gene expression or inhibition\cite{reggene}. Another kind of biological function, where multiple proteins work together, is called a ``protein complex'', where there is a high level of interaction among the involved proteins, but there is not a temporal or sequential dependency of their interactions\cite{Temp}.

Diverse technologies have made possible the compilation of Protein-Protein Interaction (PPI) networks, which contain pairs of proteins that interact in a determined experimental context. Several methods aim to discover interactions between proteins, such as yeast-two-hybrid (Y2H),
affinity purification-mass spectrometry (AP-MS) approaches, or interaction reports inferred from mining information in scientific publications\cite{Survey}. 

Despite the progress made to date, many proteins have different and multiple functions in every organism and, since biological systems are complex, there are still many knowledge gaps concerning their interactions, behavior, and functions. 
For instance, the identification of signaling pathways is a critical point to understand biological processes, as well as pathological alterations of these functions that may trigger diseases\cite{signalcancer,Cao,Gitter,Sense,Stran}.
However, prediction of biological signaling pathways from PPI networks is a complex task, mainly because PPI networks are modeled as undirected graphs and signaling pathways are intrinsically directed. 
Thus, there is a high number of possible signaling pathways to consider from a PPI network, which can produce high rates of false positives in the results. 

To date, this problem has been approached from many points of view. Gitter et al.\cite{Gitter} find pathways using a weighted PPI (to represent the confidence of the interaction) and predict alternative pathways combining random orientation and local search algorithms. Their approach first construct a weighted PPI, where weight computation is based on both the confidence in the experimental system used to detect the interaction and the number of separate research articles that report the interaction.
Their proposal aims to maximize all interaction weights in every pathway, since a pathway is more reliable if the multiplication of its weights is larger.
Cao et al.\cite{Cao} propose a pathway prediction tool that uses a distance-based metric (DSD: Diffusion State Distance), 
measuring the topological similarity of proteins in a network, adding information from databases to make it more specific 
(number of researches that prove the interactions and reference pathways). Shin et al.\cite{Shih} use a shortest path algorithm based on Dijkstra\cite{Dijkstra}, choosing the best pathways as the ones that minimize the pathway length. Jaromerska et al.\cite{distanceSim} discover signaling pathways by defining a distance measure integrating PPI network topology with Gene Ontology data and semantic similarity. 
However, this method improves Gitter's \cite{Gitter} results by a small percentage mainly in the first top-20 pathways.
 Vinayagam et al.\cite{integra} proposed a computational model that predicts activation/inhibition performing phenotype correlation among proteins and build a signed PPI network for \textit{Dropsophila melanogaster}, where the sign is positive for activation and negative for inhibition relationships. However, this approach does not predict signaling pathways. 
P-Finder~\cite{pfinder} reconstructs signaling pathways from PPI networks using gene ontology (GO) annotations and a semantic similarity metric. This method uses different algorithms based on path frequencies, network motifs, and information propagation.      
 PathLinker \cite{pathlinker} reconstructs human signaling pathways from PPI networks from receptors to transcription regulators by computing the shortest paths using a fast algorithm based on the A* heuristic. Their experimental evaluation shows that the performance of PathLinker is similar to an approach based on random walks with restarts proposed in RWR \cite{rwr}.  
 Even though some approaches integrate some biological knowledge for signaling pathway predictions, genome-scale reconstruction of signaling pathways is still challenging, mainly because causal relationships are difficult to infer\cite{integra}.

In this work, we propose a ranking algorithm that combines PPI network topology with the biological knowledge available in public databases to provide a biological context for every pathway and its interactions. Our approach is based on two steps. The first step applies an edge orientation and local search algorithm in the input PPI network to find candidate signaling pathways. The second step defines a graph model and decision algorithms that include temporal biological data, based on cell cycle dynamics, to determine which candidate pathways are biologically consistent, and then applies protein complex coverage to obtain predicted pathways.  We evaluated our method using well-known ranking metrics and found that relating biological information with PPI networks 
 significantly improves the prediction of signaling pathways.

\section{Biological knowledge}

There exists a wide range of
biological knowledge describing biological processes, components, or structures in which individual genes and proteins are known to be involved, such as protein complexes and signaling pathways. In this section, we describe what it is relevant for this work.

\subsection{Cell cycle information}
The cell cycle is a set of events where the cell grows and develops processes that lead to the duplication of its DNA and, subsequently, cell division. 
The cell cycle consists of a sequence of four phases: G1, S, G2, and M, 
where the cell components fulfill specific functions. The G1 and G2 phases are gaps, the S phase represents the synthesis (replication of its DNA), and the M phase represents the mitosis and cytokinesis. Also, there are two checkpoints, where the cell verifies if it is ready to continue with the next phase. These checkpoints are G1/S (at the end of G1) and G2/M (at the end of G2)\cite{Sync}.

\begin{figure}[ht]
\begin{tabular}{ll}
\hspace{-1.3cm}
\includegraphics[scale=0.5]{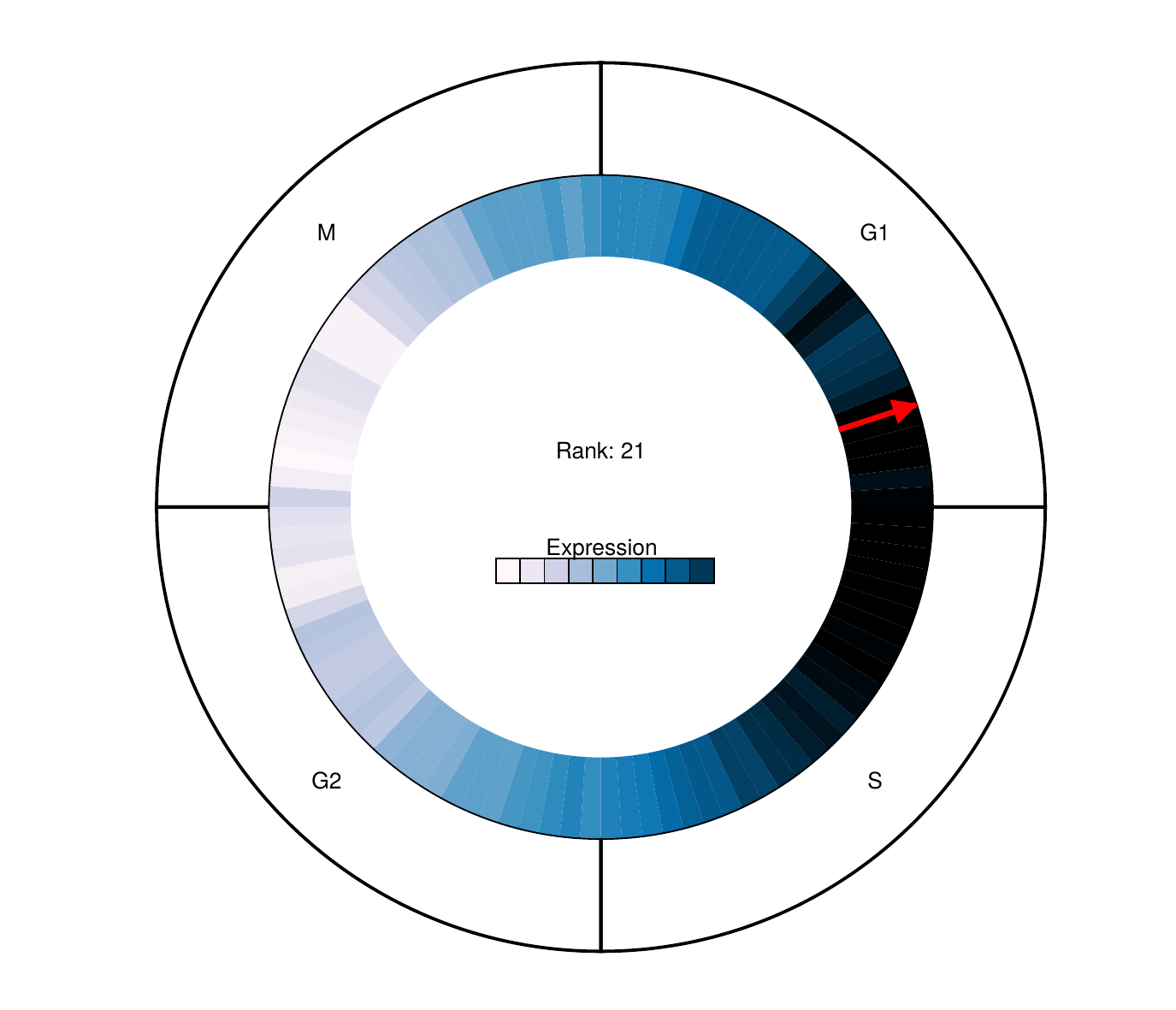}
&
\hspace{-1.2cm}
\includegraphics[scale=0.5]{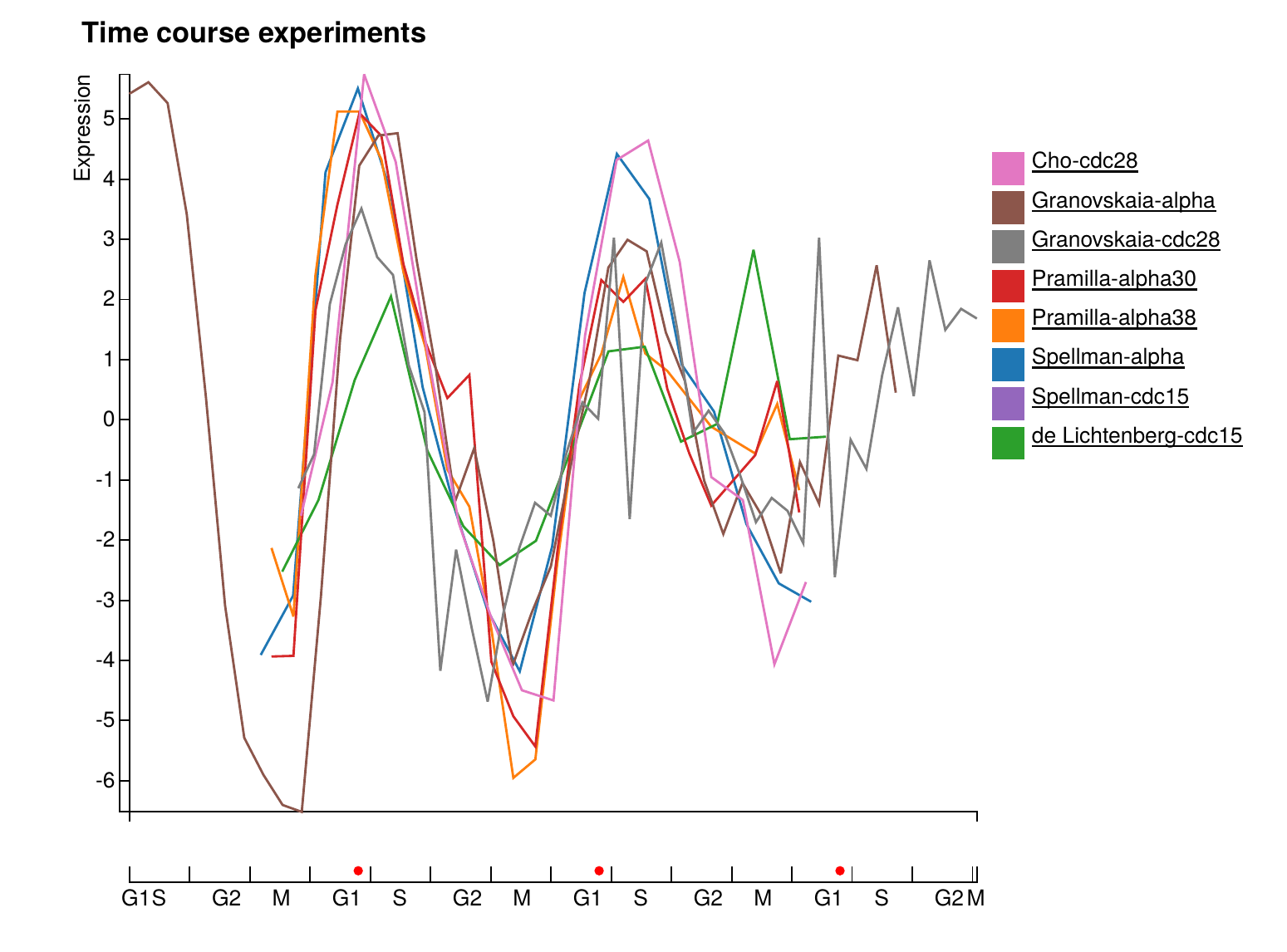}

\end{tabular}
\caption{Capture from Cyclebase 3.0. Information available for protein YBR088C with including temporal expression data from eight experiments.}
\label{fig1ab}
\end{figure}

There are many available datasets related to the mitotic cell cycle. These datasets include microarray-based time courses of mRNA expression, mass-spectrometry-based proteomics on protein expression during the cell cycle, systematic screens for cyclin-dependent kinase (CDK) substrates, and high-content screening for knockdown phenotypes. All these datasets provide detailed information about the mitotic cell cycle and its many regulatory layers. As combining and analyzing all this information is a complex task, Cyclebase \cite{Cyclebase} aims to address this problem by processing different experimental data, mapping common gene identifiers and normalizing experiments onto a common timescale, facilitating direct comparison of expression profiles between different experiments related to an organism. Version 3.0 of Cyclebase includes new mRNA and protein expression data, and integrates cell cycle phenotype information from high-content screens and model-organism databases.
Cyclebase also provides an easy way of obtaining information about cell cycle peak gene expression and phenotypes of individual genes.
Figure \ref{fig1ab} shows the information available for a gene (YBR088C) in yeast (\textit{Saccharomyces cerevisiae}), where the gene expression in different phases of the cell cycle and the time course experiments are displayed, and the periodic behavior of gene expression can be observed ( in red arrow in Figure~\ref{fig1ab}-left and red dot in Figure~\ref{fig1ab}-right). In addition we show a sample of the information displayed for several genes related to peak expression and phenotype for a set of yeast genes, including YBR088C, first row in Table \ref{fig1c}.

 \begin{table}[ht] 
 \begin{tabular}{lllll}
\hline
Primary name&Type&Identifier&Peaktime&Phenotypes\\
 \hline
POL30&Saccharomyces cerevisiae&YBR088C&G1&\\
CLN2&Saccharomyces cerevisiae&YPL256C&G1&G1\\
MRC1&Saccharomyces cerevisiae&YCL061C&G1&G1/S,S\\
 \hline
\end{tabular}
 \caption{Set of genes with corresponding peak expression phase and phenotype}
 \label{fig1c}
 \end{table}

\subsection{Protein complex information}
Proteins are known to participate in several biological processes such as transport, signaling, and metabolic and enzymatic catalysis.
Most proteins interact with others forming functional units, called protein complexes, which allows them to perform biological functions collaboratively. Many proteins participate in different protein complexes according to the functional needs of the organism.
Understanding the functions of proteins is important for many diseases since some research studies have shown that the deletion of some proteins in a network may have lethal effects on organisms\cite{Lethal}. This has been an important motivation for the research community to propose different prediction methods for protein functions and protein complexes\cite{Hernandez,Marco,Survey}. Moreover, there are gold standards with curated protein complexes for different organisms. In the case of yeast, one of the most used gold standards is CYC2008~\cite{CYC}, which contains 408 manually curated heteromeric protein complexes and has been used as a reference by many protein complex prediction tools~\cite{Survey2}.

\section{Method}

In this section, we formally define the signaling pathway discovery problem and propose a method that incorporates biological data to infer causal or temporal relationship in the
signal flow. 

\subsection{Problem formulation}

Let $G(V,E,w)$ be a weighted undirected graph, which models a PPI network with proteins as vertices in $V$, protein interactions as edges in $E$, and weights, $w$, as the confidence of a real protein interaction. We denote $w(e)$ as the weight (confidence) of an edge $e$ in $E$. 
Let also assume a set of pairs $(s_i,t_i)$ of $source$, $target$ proteins in $V$ and a maximum path length $h$ in $G$ for all pairs $(s_i,t_i)$. 

We formulate the problem of discovering signaling pathways in two steps. The first step has the goal of defining an edge orientation. For the first step, we use Gitter et al.\cite{Gitter} to find \textit{candidate pathways}. Gitter et al.\cite{Gitter} formulate the problem for length-bounded edge orientation pathways in weighted interaction networks and show that it is NP-hard. 
The authors consider this problem as Maximun Edge Orientation (MEO) and propose an heuristic based on random edge orientation with a local search algorithm, which objective function is to maximize all the weights in the pathway, where the weights represent the protein interaction confidences.  

The second problem is defined as a graph model and decision rules, where each candidate pathway is evaluated using biological data that include temporal dynamics using cell cycle and protein complex information. As a result, we obtain a set of signaling pathways, which we call \textit{consistent pathways}. Afterwards, we verify each consistent pathway for protein complex coverage to obtain \textit{predicted pathways}. 

\begin{definition}{Candidate pathway.}\\
A candidate pathway is a path $P = <v_1, v_2, \dots , v_m>$, where each pair of consecutive vertices in a path forms an edge $(v_j,v_{j+1}) \in E$, in which the first vertex is a source, $s_i$, and the last one is its corresponding target, $t_i$.
\end{definition}

\begin{definition}{Consistent pathway.}\\
A consistent pathway is a candidate pathway where all of its edges, when interpreted as interactions between proteins, satisfy temporal cell cycle dynamics or protein complex rules (Algorithm \ref{Alg1}).
\end{definition}

\begin{definition}{Protein Complex.}\\
A protein complex is a set $pc = \{p_1, p_2, \dots , p_n\}$, where $p_i$ is a protein.
\end{definition}

\begin{definition}{Predicted pathway.}\\
A predicted pathway is a consistent pathway where all of its nodes, when interpreted as proteins, satisfy the given ratio of protein complex coverage. Protein complex coverage is defined as the ratio of the maximum number of proteins in a pathway that belong to a single protein complex with respect to the total number of proteins in a pathway.
\end{definition}

\subsection{Approach}

We define a cell cycle graph $CG=(A,B)$, where $A$ represents the cell cycle phases and $B$ transitions between them. Further, $N(x)$ returns the transitions from one phase to others in $CG$. 

 \begin{equation}\label{rules}
  		   A = \{ G1, G1/S, S, G2, G2/M, M \}
\end{equation}

 \begin{equation} 		   
         B = \{ (x,y) \in A \times A \}
\end{equation}
         
\begin{equation}         
  		N(x) = \{ y \in A / (x,y) \in B \} 
\end{equation}


The idea is based on the fact that, if two consecutive proteins in a candidate pathway are expressed in the same cell cycle phase, which is a self-loop edge in the graph, or if the second protein is expressed in a phase that follows the one of the first protein, then the interaction is more likely to be real. 
As the information extracted from Cyclebase includes two checkpoints, G1/S and G2/M, we use them as a part of the cycle, so that the sequence remains in order. 
We based our method on the information reported by Cyclebase, which already estimates peak gene expression by a probabilistic approach using several gene expression experiments.

\begin{figure}[ht]
 \centering
\includegraphics[scale=1.2]{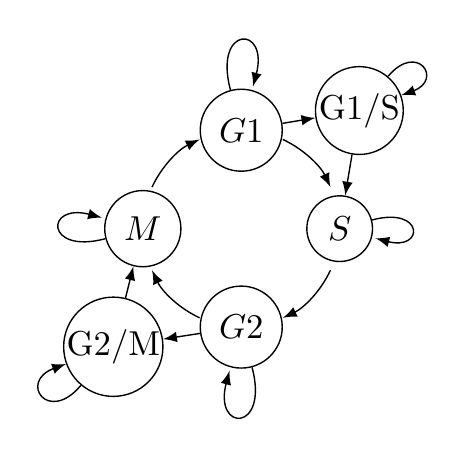}
 \caption{Graph representation of a cell cycle $CG(A,B)$.}
 \label{fig2}
\end{figure}

We incorporate temporality in the form of cell cycle phases to check if two consecutive proteins in a candidate pathway are likely to participate in a pathway.
If all proteins forming edges in a candidate pathway satisfy the expected transitions in cell cycle dynamics or protein complex involvement, the \textit{candidate pathway} is a \textit{consistent pathway}.

 We model the cell cycle transition as the function $T$ that maps each protein (seen as a node in $V$) to the set of cell cycle phases (seen as nodes in $A$) where it is expressed. To define $T$, we use information from Cyclebase 3.0 \footnote{www.cyclebase.org}, from where we may use the information on cell cycle peak expression (in which case the resulting function is called $T_{pk}$), or the information on cell cycle phenotype (in which case we call $T_{ph}$ the resulting function). In addition, we consider the protein complex involvement function $L$ to map proteins to protein complexes in which they participate, where $PC$ are predicted or gold standard complexes. 
The procedure that involves functions $T$ and $L$ can be used in any order as they are independent.

\bigskip

\begin{tabular}{ll}

$T: V \mapsto C \subseteq A$
&
\hspace{1cm}
$L: V \mapsto R \subseteq PC$
\end{tabular}

\bigskip

Then, given $CG(A,B)$ and $\mathcal{P} = \{P_1, P_2, \dots , P_n\}$ , we decide whether a candidate pathway $P_i$ is TRUE or FALSE as shown in Algorithm \ref{Alg1}. Consistent pathways are candidate pathways that are TRUE after applying Algorithm \ref{Alg1}.

\bigskip

\begin{algorithm}
\caption{Signaling consistent pathway decision using cell-cycle and protein complex rules.}
\label{Alg1}
\begin{algorithmic}[1]
\REQUIRE Candidate pathway $P \in \mathcal{P}$, functions $T_{pk}$, $T_{ph}$, $L$. 
\ENSURE Returns TRUE (consistent pathway) if a candidate pathway satisfies the cell-cycle dynamics or protein complex rules.
\FOR {$v_i,v_{i+1} \in P$}
\IF {$(|T_{pk}(v_{i+1}) \cap N(T_{pk}(v_{i}))|>0) \vee (|T_{ph}(v_{i+1}) \cap N(T_{pk}(v_{i}))|>0) \vee (|T_{pk}(v_{i+1}) \cap N(T_{ph}(v_{i}))|>0) \vee (|T_{ph}(v_{i+1}) \cap N(T_{ph}(v_{i}))|>0)$}
\STATE continue
\ELSIF {($|L(v_i) \cap L(v_{i+1})|>0$)}
\STATE continue
\ELSE 
\RETURN FALSE
\ENDIF
\ENDFOR
\RETURN TRUE
\end{algorithmic}
\end{algorithm}

The first part of the Algorithm \ref{Alg1} (Cell Cycle rule), lines 2-3, evaluates whether a protein-protein interaction, $(v_{i},v_{i+1})$ is supported by: 
\begin{enumerate}
 \item Peaks of expression: where $v_{i+1}$ peak must be in the next phase of $v_{i}$ peak.
 \item Peaks and phenotypes: where $v_{i+1}$ phenotype must be in the next phase of $v_{i}$.
 \item Phenotypes and peaks: where $v_{i+1}$ peak must be in the next phase of $v_{i}$ phenotype.
 \item Phenotypes: where $v_{i+1}$ phenotype must be in the next phase of $v_{i}$ phenotype.
\end{enumerate}

The next phase in the graph of a protein can be only as indicated in Figure \ref{fig2}, that is, staying in the same one or going to other phases in the Cyclebase cell cycle. In case the cell cycle is not satisfied we verify the complex rule, line 4, that is, our approach verifies if either the cell cycle rule or the complex rule is satisfied. If none of these two rules is satisfied, we discard such interaction (lines 6-7) and, consequently, the candidate pathway.

 The pathways that pass the filter of Algorithm \ref{Alg1} are furthered filtered as follows. Taking advantage of the high relationship between the proteins when they collaborate in a complex, we evaluate all the proteins of each \textit{consistent pathway} to calculate the highest coverage that the complexes have on the pathways. This can be seen in Algorithm \ref{Alg2}, where we  
compute the number of common protein complexes where proteins in a \textit{consistent pathway} participate in (lines 1-3 in Algorithm \ref{Alg2}).
Then, we compute the protein complex coverage $r$ (line 4 in Algorithm \ref{Alg2}) and choose the pathways that are over the threshold given by the coverage ratio $cv$.
This filter is related with the level of cohesion within the pathway, appreciating that certain proteins in it carry out functions together, beyond temporality.

\begin{algorithm}
\caption{Signaling consistent pathway with protein complex coverage.}
\label{Alg2}
\begin{algorithmic}[1]
\REQUIRE Consistent pathway $P_c$ and coverage ratio $cv$. 
\ENSURE Returns TRUE (Predicted pathway) if a consistent pathway satisfies the $cv$.

\FOR {$v_i \in  P_c$}
\STATE $Freq[L(v_i)]++$
\ENDFOR
\STATE $r=max(Freq)/|P_c|$
\IF {$r>=cv$}
\RETURN TRUE
\ELSE 
\RETURN FALSE
\ENDIF

\end{algorithmic}
\end{algorithm}

\subsection{Ranking algorithms}\label{subsec_rankalgs}

A post-processing step of our method ranks consistent pathways according to different criteria. In this section, we describe the main metrics used in such process, which are the edge and path metrics shown in Equation \ref{edgemetrics}. These metrics compute a value for each pathway $P_p$, which is used to rank the pathways in a priority queue.

\begin{equation}\label{edgemetrics}
 Edge\; Weight = \min/\max/Avg\;_{\forall_e \in P_p} w(e) 
\end{equation}
\begin{equation}
 Edge\; Use = \min/\max/Avg\;_{\forall_e \in P_p} Freq(e) 
\end{equation}
\begin{equation}
 Path\; Weigh = \prod_{e \in P_p} w(e)
\end{equation}


The \textit{Edge Weight} returns the minimum, average or maximum weight of an interaction in a pathway. The \textit{Edge Use} is the number of times a certain edge is present in all predicted pathways. It is also considered as in its minimum, maximum, and average values. The \textit{Path Weight} is the multiplication of all the edge weight values, $w(e)$, from a pathway. Path weight is used to break ties when ranking by other metrics.

In preliminary experiments, we also considered vertex centrality metrics such as \textit{degree}, \textit{betweenness} and \textit{closeness}. However, such results are not reported here since edge and path metrics always outperformed vertex metrics in our evaluation.

\section{Evaluation}

This section describes the experimental evaluation performed to compare the proposed method with state of the art methods for discovering signaling pathways. We first describe the experimental setup, which consists of input datasets, gold standards, evaluation metrics and alternative methods used for comparison, and then the results and a discussion about them.

\subsection{Experimental setup}

We replicate the experimental setup used by Gitter et al.\cite{Gitter}. Hence, we use their PPI network, Gold Standard and reference pathways. Next, we briefly explain each of them. The PPI network consists of highly plausible interactions driven by the union and analysis of different PPI databases such as MINT, BioGrid, and IntAct. Gitter et al.\cite{Gitter} builds this PPI using both confidence of experimental system used to detect the interactions and research articles supporting the interactions between proteins.
This PPI contains 3,446 proteins and 10,944 interactions. 

The set of reference pathways consists of real pathways extracted from KEGG and Science Signaling Database of Cell Signaling. As they stated, signaling pathways from KEGG (MAPK signaling pathway) and the Science Signaling Database of Cell Signaling (Pheromone pathway and High Osmolarity Glycerol, HOG pathway) contain an average of 5 edges between a source and its closest target. We have included the KEGG Cell wall stress pathway.
Gitter et al.\cite{Gitter} made this observation to define $h=5$ as the length of the pathway in their experiments, and then we use the same value for $h$.
This value was calculated by getting the shortest path from any source to each target, with a PPI network only with interactions from each of the four pathways in evaluation (MAPK, Pheromone, Cell wall stress, and HOG).  This study is useful to bound the length of the reference pathways and the predicted pathways, since the longer the pathway, the more computational resources (memory and time) are needed.

In all gold standard pathways, we discarded inhibition interactions and only considered activation interactions, because inhibition interactions lead to stopping in the cascade of interactions. In the reference PPI from KEGG and Science Signaling Database of Cell Signaling (where the gold standard is made from), there were only four inhibition interactions.

Reference pathways were generated from the list of 16 sources and 16 targets chosen by Gitter et al.\cite{Gitter}, as a list of vertices without a parent vertex (in the case of sources) and a list of vertices without children (in the case of targets).
In addition, we added the Cell wall stress pathway adding three sources and three targets, having a total of 19 sources and 19 targets. All sources and targets are displayed in Table \ref{syt}. These sources and targets were used to generate the candidate pathways in the initial step.  

\begin{table}
\caption{List of sources and targets used.}

\begin{tabular}{p{0.21\textwidth} p{0.22\textwidth} p{0.21\textwidth} p{0.22\textwidth}}
\hline
\multicolumn{2}{c}{Source}&\multicolumn{2}{c}{Target}\\
Standard name&Systematic name&Standard name&Systematic name\\
 \hline
SLN1&YIL147C&CDC42&YLR229C\\
YCK1&YHR135C&HOG1&YLR113W\\
YCK2&YNL154C&STE7&YDL159W\\
SHO1&YER118C&STE20&YHL007C\\
MF(ALPHA)2&YGL089C&DIG2&YDR480W\\
MID2&YLR332W&DIG1&YPL049C\\
RAS2&YNL098C&PBS2&YJL128C\\
GPR1&YDL035C&FUS3&YBL016W\\
BCY1&YIL033C&STE5&YDR103W\\
STE50&YCL032W&GPA1&YHR005C\\
MSB2&YGR014W&MSN1&YOL116W\\
SIN3&YOL004W&FKS2&YGR032W\\
RGA1&YOR127W&FUS1&YCL027W\\
RGA2&YDR379W&STE12&YHR084W\\
ARR4&YDL100C&SWI4&YER111C\\
MF(ALPHA)1&YPL187W&FLO11&YIR019C\\
PKH1 & YDR490C & RLM1 & YPL089C\\
PKH2 & YOL100W & SWI4 & YER111C\\
PKH3 & YDR466W & SWI6 & YLR182W \\
 
 \hline
\end{tabular}
\label{syt}
\end{table}

Also, we included biological datasets that register the peak expression data and phenotypes available in Cyclebase\cite{Cyclebase}. We consider predicted yeast protein complexes obtained by Hernandez et al. \cite{Hernandez} and yeast protein complex gold standard CYC2008\cite{CYC}. We used a ranking scheme choosing the top-$k$ predicted pathways that considers the different pathway metrics described in Section~\ref{subsec_rankalgs}.

We used as parameters: $h$, the length of the pathways to consider; $cv$, intended to consider different ratios of protein complex coverage; and $k$, the size of the top-$k$ rankings. We compared our method using different configurations, as shown in Table \ref{configs}, which includes Gitter's method that is used as a baseline. We also compare our results with PathLinker \cite{pathlinker} and RWR \cite{rwr}. Both implementations are available at \url{http://bioinformatics.cs.vt.edu/\~murali/supplements/2016-sys-bio-applications-pathlinker/}. 

\begin{table}
\caption{Method configurations.}

\begin{tabular}{p{0.18\textwidth}p{0.75\textwidth}}
\hline
Config & Description\\
\hline
 G& Gitter's method, using the best values from 10 tests (as it is a random model, results may vary).\\
 G-cv & Gitter's method plus protein complex coverage (i.e. using candidate pathways as input and protein complex coverage in Algorithm \ref{Alg2}). We use protein complex coverage ratio \textit{cv} in percentage. \\

 GC& Gitter for candidate pathways plus applying Cell cycle rule (defined in Algorithm \ref{Alg1}).\\

 GCC& Gitter for candidate pathways, applying Cell cycle rule and protein complex rule (defined in Algorithm \ref{Alg1}).\\

 GCC-cv& Gitter for candidate pathways, applying Cell cycle rule, protein complex rule (defined in Algorithm \ref{Alg1}), and protein complex coverage (defined in Algorithm \ref{Alg2}).\\
 
 PathLinker & Pathways predicted by PathLinker. \\
 PathLinker-RWR & Pathways predicted by PathLinker RWR implementation.\\
 \hline
\end{tabular}
\label{configs}
\end{table}

\subsection{Evaluation metrics}

We follow the definitions of complete and partial match as given in Gitter et al.\cite{Gitter} to define true positives.
Hence, a predicted pathway is considered as a true positive if the pathway has at least three of the five interactions consecutively and it is found in at least one reference pathway. 

Taking into account this definition, we considered the typical measures used to evaluate effectiveness of ranked retrieval\cite{IR}, such as \textit{Precision@k}, \textit{Recall@k}, $F_1Score@k$, and MAP. \textit{Precision@k} evaluates the effectiveness of a \textit{top-k} prediction list as the ratio of the number of true positives the list contains and the list size. In other words, it gives an idea about the portion of positive results in the retrieved list. An important property of this measure is that it tends to decrease as $k$ increases. On the other hand, \textit{Recall@k} is computed as the ratio of the number of true positives the list contains and the number of positive samples in the ground truth. It gives an idea about the coverage of truth samples obtained from the predicted ranking. By definition, it monotonically increases with $k$. Values are computed as described in Equations \ref{evalmetrics}. In these equations, $I_k(u)$ represents the list of pathways returned by a method and $I_u^+$ represents positive samples in the ground truth. Hence, $I_u^+ \cap I_k(u)$ are the true positives. $F_1Score@k$ is the harmonic average of \textit{Precision@k} and \textit{Recall@k}. In the next section, we will show recall-precision curves, which provide a graphical way of examining the trade-off between these two measures.

\begin{equation}\label{evalmetrics}
  Precision@k = \frac{| I_u^+ \cap I_k(u)|}{|I_k(u)|} 
\end{equation}

\begin{equation}
 Recall@k = \frac{| I_u^+ \cap I_k(u)|}{| I_u^+ |} 
\end{equation}

\begin{equation}
F_1Score@k = \frac{2 \times Precision@k \times Recall@k}{Precision@k + Recall@k}
\end{equation}


The last measure we consider, MAP, is based on the \emph{average precision}, which may be interpreted as the area under the recall-precision curves. Hence, it aggregates all the recall values in a single measure, which may simplify the comparison between different methods.
The Mean Average Precision or MAP is just the arithmetic mean over a set of predictions of the \emph{average precision} values. 

In order to correctly interpret the results, it is important to notice that, unlike other domains in which these measures have been used, in this domain the ground truth is incomplete. In plain words, true positives represent pathways that have been shown to exist by biological experiments, but false positives do not necessarily represent pathways that do not actually exist. Hence, low precision values are not as important as in other domains where the ground truth is complete.

\subsection{Results}
As mentioned in the previous section, we used a high confidence PPI network, signaling pathway gold standards, path ranking metrics, and path length of 5 interactions as defined in Gitter approach~\cite{Gitter}.
We considered all configurations described in Table \ref{configs}, using $cv=10,20,30,40,50$\% for all path ranking metrics defined in Equations \ref{edgemetrics}.
 
We first evaluate different configurations of our method and compare them with the baseline~\cite{Gitter}. To do that we measure the number of true positives in the top-100 results obtained by sorting the pathway predictions using the defined path ranking metrics. For protein complex coverage we use predicted protein complexes from the DAPG method~\cite{Hernandez} and from the gold standard CYC2008~\cite{CYC}. Table~\ref{table_edgepred} shows the number of true positives out of the top-100 results obtained using predicted protein complexes and Table~\ref{table_edge} using gold standard protein complexes. As observed, both tables show similar results.
Furthermore, we observe that using cell cycle and protein rules, as well as protein complex coverage, are better alternatives than using only the random edge orientation with local search proposed by Gitter et al.\cite{Gitter}, since the number of true positives increases about 27\% (ratio between our method GCC-30\% and G-30\%).

\begin{table*}[htp!]
\small
\caption{Ranking results (path and edge measures) from the top-$100$ consistent signaling pathways using predicted protein complexes. Values in bold show best results.}

\hspace{1.6cm}
 \begin{tabular}{lp{0.08\textwidth}p{0.08\textwidth}p{0.08\textwidth}p{0.08\textwidth}p{0.08\textwidth}p{0.08\textwidth}p{0.08\textwidth}}
 \hline
  Method & Path Weight & Max Edge Weight & Avg Edge Weight & Min Edge Weight & Max Edge Use & Avg Edge Use & Min Edge Use \\
  \hline
  G & 31 & 7 & 31 & 34 & 0 & 0 & 0 \\
  G-10 & 31 & 7 & 31 & 34 & 0 & 0 & 0 \\
  G-20 & 31 & 7 & 31 & 32 & 0 & 0 & 0\\
  G-30 & 31 & 7 & 31 & 31 & 0 & 0 & 8\\
  G-40 & 26 & 3 & 26 & 25 & 0 & 0 & 0\\
  G-50 & 26 & 3 & 26 & 25 & 0 & 0 & 0\\
   GC& 41 & 12 & 43 & 36 & 7 & 4 & 3\\
   GCC& \bf{45} & 17 & 44 & 42 & 6 & 2 & 0\\
   GCC-10& \bf{45} & 17 & 44 & 42 & 6 & 2 & 0\\
   GCC-20& \bf{47} & 17 & \bf{45} & 44 & 6 & 2 & 0\\
   GCC-30& \bf{47} & 17 & \bf{45} & 44 & 6 & 2 & 0\\
   GCC-40& 40 & 10 & 43 & 41 & 0 & 1 & 0\\
   GCC-50& 42 & 10 & 43 & 41 &0  &1  &0 \\
  \hline
  
 \end{tabular}
\label{table_edgepred}
\end{table*}

\begin{table*}[htp!]
\small

\caption{Ranking results (path and edge measures) from the top-$100$ consistent signaling pathways using protein complexes in gold standard CYC2008. Values in bold show best results.}

\hspace{1.6cm}
 \begin{tabular}{lp{0.08\textwidth}p{0.08\textwidth}p{0.08\textwidth}p{0.08\textwidth}p{0.08\textwidth}p{0.08\textwidth}p{0.08\textwidth}}
   \hline
  Method & Path Weight & Max Edge Weight & Avg Edge Weight & Min Edge Weight & Max Edge Use & Avg Edge Use & Min Edge Use \\
  \hline
  G & 31 & 7 & 31 & 34 & 0 & 0 & 0 \\
  G-10 & 31 & 7 & 31 & 34 & 0 & 0 & 0 \\
  G-20 & 32 & 7 & 32 & 37 & 0 & 1 & 8\\
  G-30 & 32 & 7 & 32 & 37 & 0 & 0 & 8\\
  G-40 & 28 & 3 & 28 & 26 & 0 & 0 & 0\\
  G-50 & 28 & 3 & 28 & 26 & 0 & 0 & 0\\
   GC& 41 & 12 & 43 & 36 & 7 & 4 & 3\\
   GCC& \bf{45} & 17 & 44 & 42 & 6 & 2 & 0\\
   GCC-10& \bf{45} & 17 & 44 & 42 & 6 & 2 & 0\\
   GCC-20& \bf{47} & 17 & \bf{45} & \bf{46} & 0 & 1 & 5\\
   GCC-30& \bf{47} & 17 & \bf{45} & \bf{46} & 0 & 1 & 5\\
   GCC-40& 40 & 10 & 42 & 41 & 0 & 0 & 0\\
   GCC-50& 40 & 10 & 42 & 41 &0  &0  &0 \\
  \hline
  
 \end{tabular}
\label{table_edge}
\end{table*}

\begin{figure}[ht]
\hspace{-5mm}
\begin{tabular}{ll}
\hspace{-1.2cm}
\includegraphics[scale=0.27]{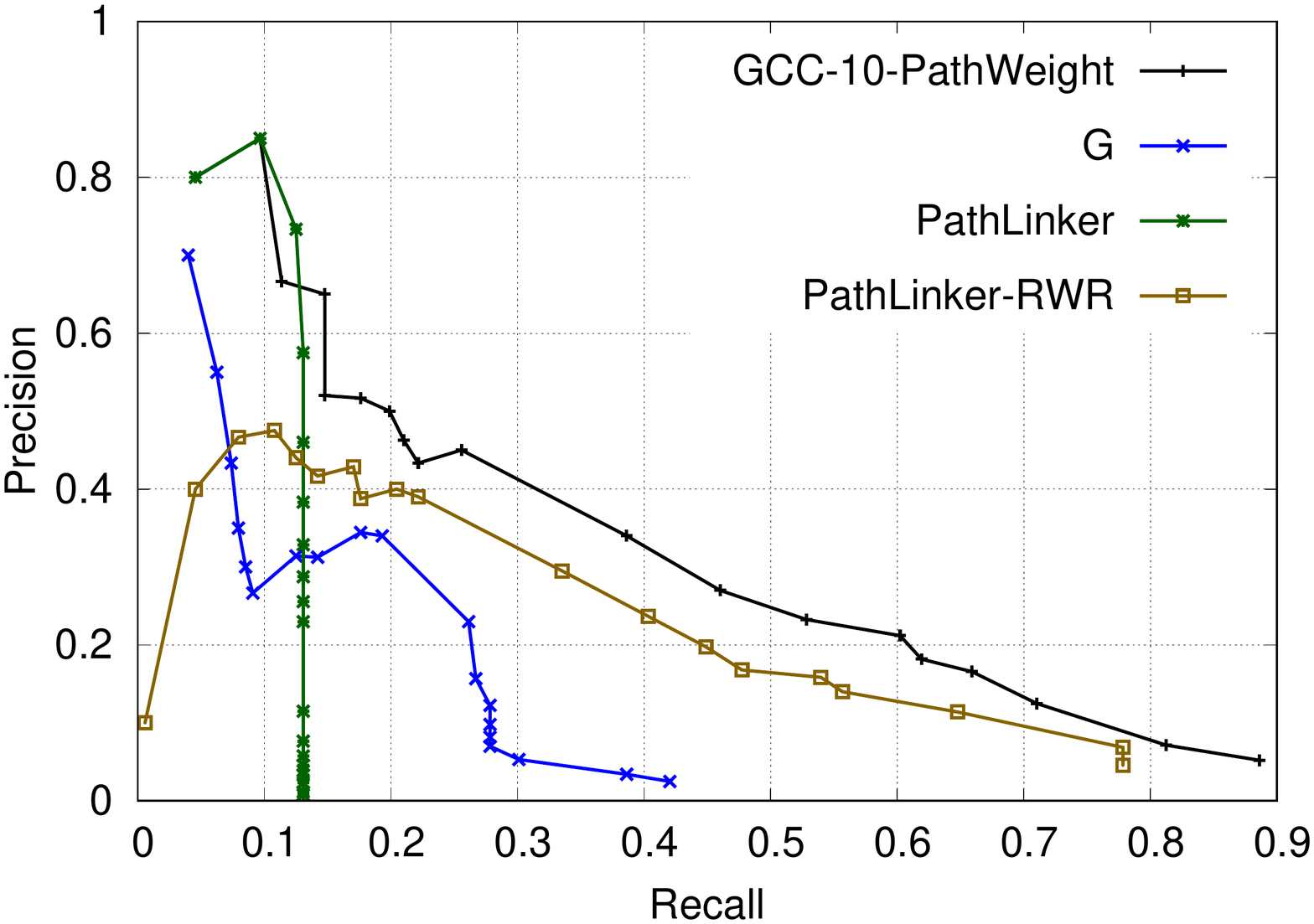}
&
\hspace{-1.6cm}
\includegraphics[scale=0.27]{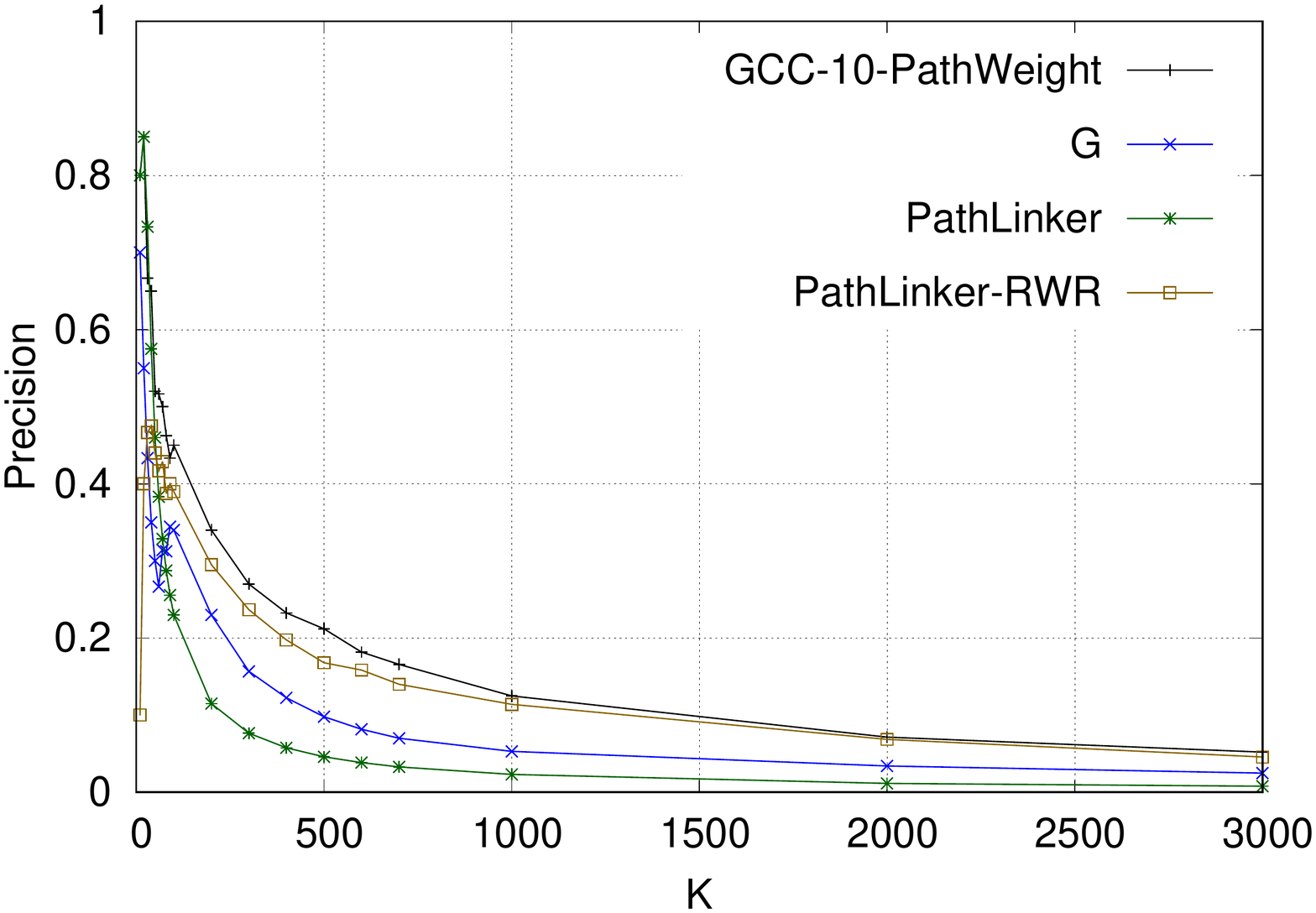}
\\
\vspace{-1cm}
\hspace{-1.2cm}
\includegraphics[scale=0.27]{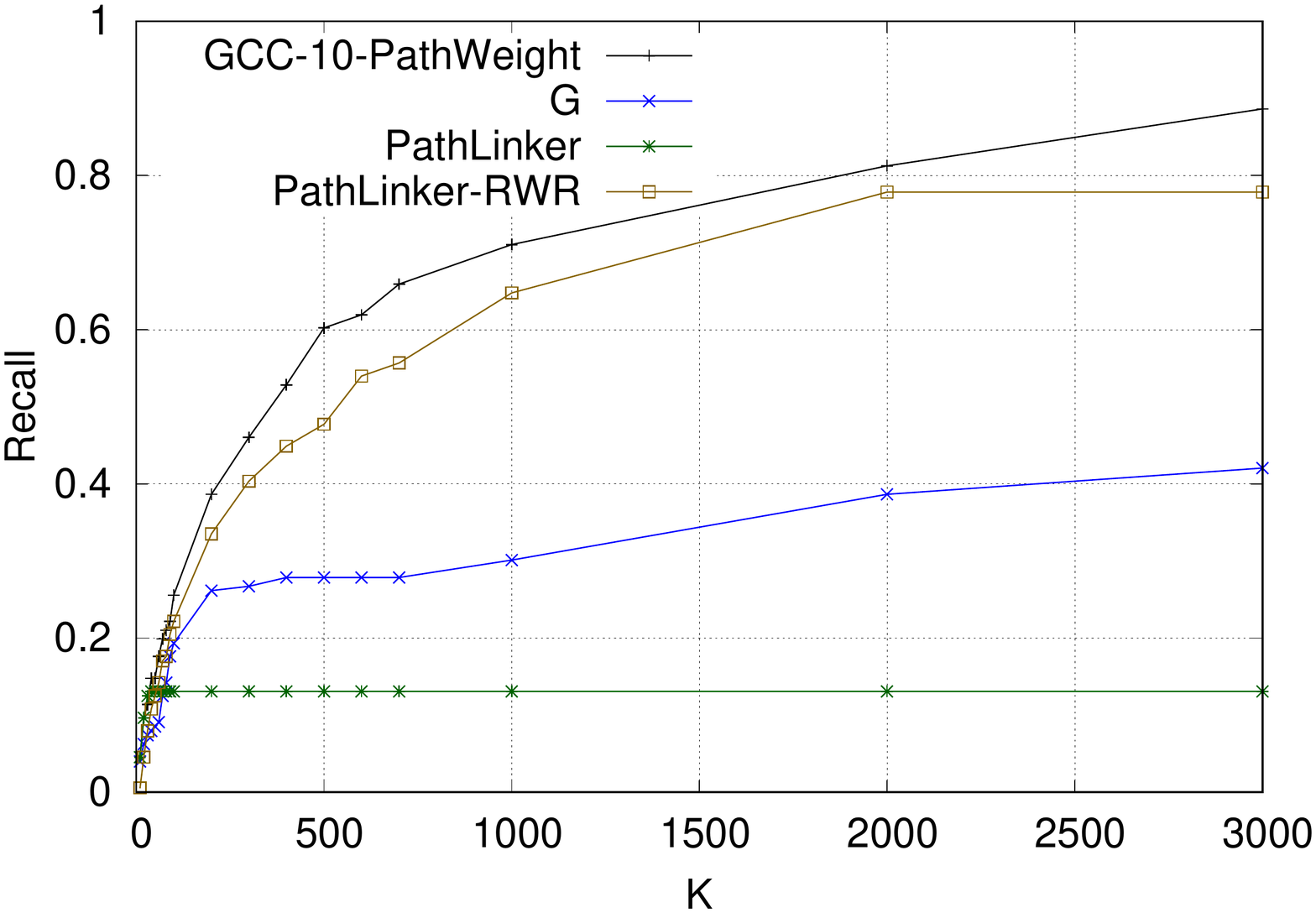}
&
\hspace{-1.4cm}
\includegraphics[scale=0.27]{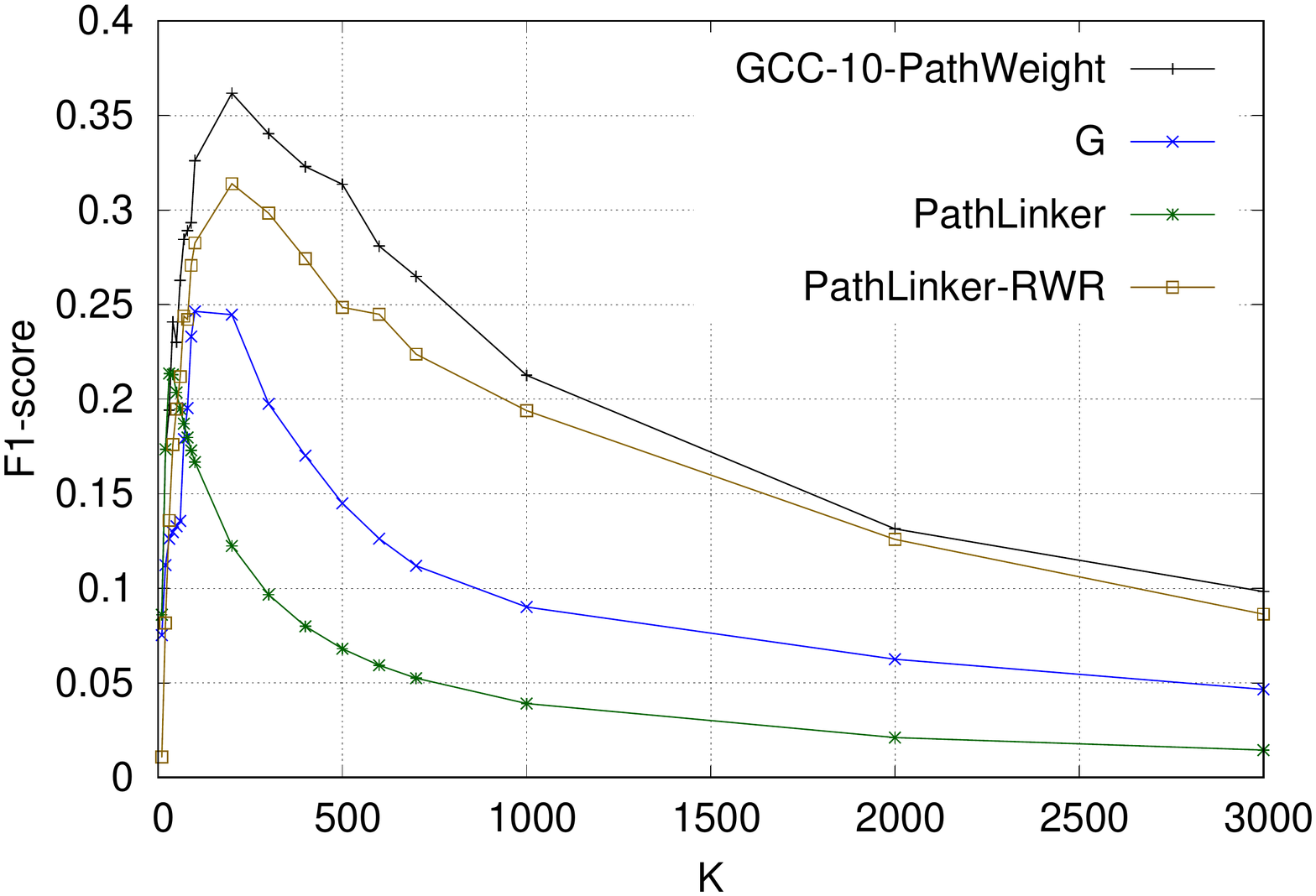}
\\
\end{tabular}
\caption{Precision/Recall $@k$, Precision, Recall and $F_1Score$ for different $k$ values. Comparison among Gitter method (G) using best path metrics and G with best protein complex coverage, our method using best cell cycle and protein complex coverage (GCC) settings, PathLinker and PathLinker-RWR best results. Best configuration obtained with metric PathWeight using protein coverage of 10\% (GCC-10).}

\label{best}
\end{figure}

We promote our best configurations and compare them with state-of-the-art methods using standard measures: precision, recall, $F_1Score@k$ and MAP (Mean Average Precision). We summarize the results in Figure~\ref{best}.
The top left graph shows the best results
including the original results of G, GCC, 
PathLinker and PathLinker-RWR.
Figure~\ref{best}-left shows that our approach provides the best performance regarding $precision$ and $recall@k$, achieving better precision for the whole recall spectrum.
The other graphs in Figure~\ref{best} show the $precision@k$, $recall@k$ and $F_1Score@k$. We observed that our method provides the best results for all metrics. We also observed that the second best method is PathLinker-RWR, whereas PathLinker does not behave well. We tested PathLinker increasing $k$, but results do not improve. 
Also, we compare the overall performance using MAP with the best configurations for G, PathLinker-RWR and GCC. We find that, with GCC, MAP is 0.66 whereas with G it is 0.45 and with PathLinker-RWR it is 0.47, which gives us an improvement of 46\% in both cases.

In addition, we include a biological metric that measures the biological significance of predicted
signaling pathways using enrichment analysis. We defined biological significance as the ratio of significant pathways with respect the top-k predicted pathways. This computation is similar to the one described for protein complexes \cite{go}. 
We use gprofiler \cite{gprofiler}, a recent tool that automatically considers the latest gene ontology and annotations. We considered its python  client application using the complete predicted pathway protein gene list, excluding electronic GO annotations and using \textit{p-value} smaller than 0.01 with minimum number of genes as the gene list size. This tool is publicly available at \url{http://biit.cs.ut.ee/gprofiler/}.   
Table \ref{enrich} shows these results for the top-$100$ ranked pathways for the best methods.  

\begin{table}
\caption{Biological significance for predicted pathways with the best methods, GCC-10 and PathLinker-RWR.}

\begin{tabular}{p{7mm}|c|c|c|c|c|c|c|c|c|c|}
\cline{2-11}
& \multicolumn{10}{ |c| }{k} \\ \cline{1-11}
\multicolumn{1}{ |c|  }{Method}& 10 & 20 & 30 & 40 & 50 & 60 & 70 & 80 & 90 &100 \\ \cline{1-11}
\multicolumn{1}{ |c|  }{GCC-10} & 1.00 & 1.00 & 0.94 & 0.83 & 0.84 & 0.85 & 0.88 & 0.88 & 0.84 & 0.81 \\\cline{1-11}
\multicolumn{1}{ |p{1.6cm}|  }{PathLinker-RWR} & 0.90& 0.88& 0.81& 0.78& 0.82& 0.73& 0.74& 0.76& 0.77 & 0.71\\
\hline
\end{tabular}
\label{enrich}
\end{table}

\begin{figure}[ht]
\hspace{-5mm}
\begin{tabular}{ll}
\hspace{-1.2cm}
\includegraphics[scale=0.27]{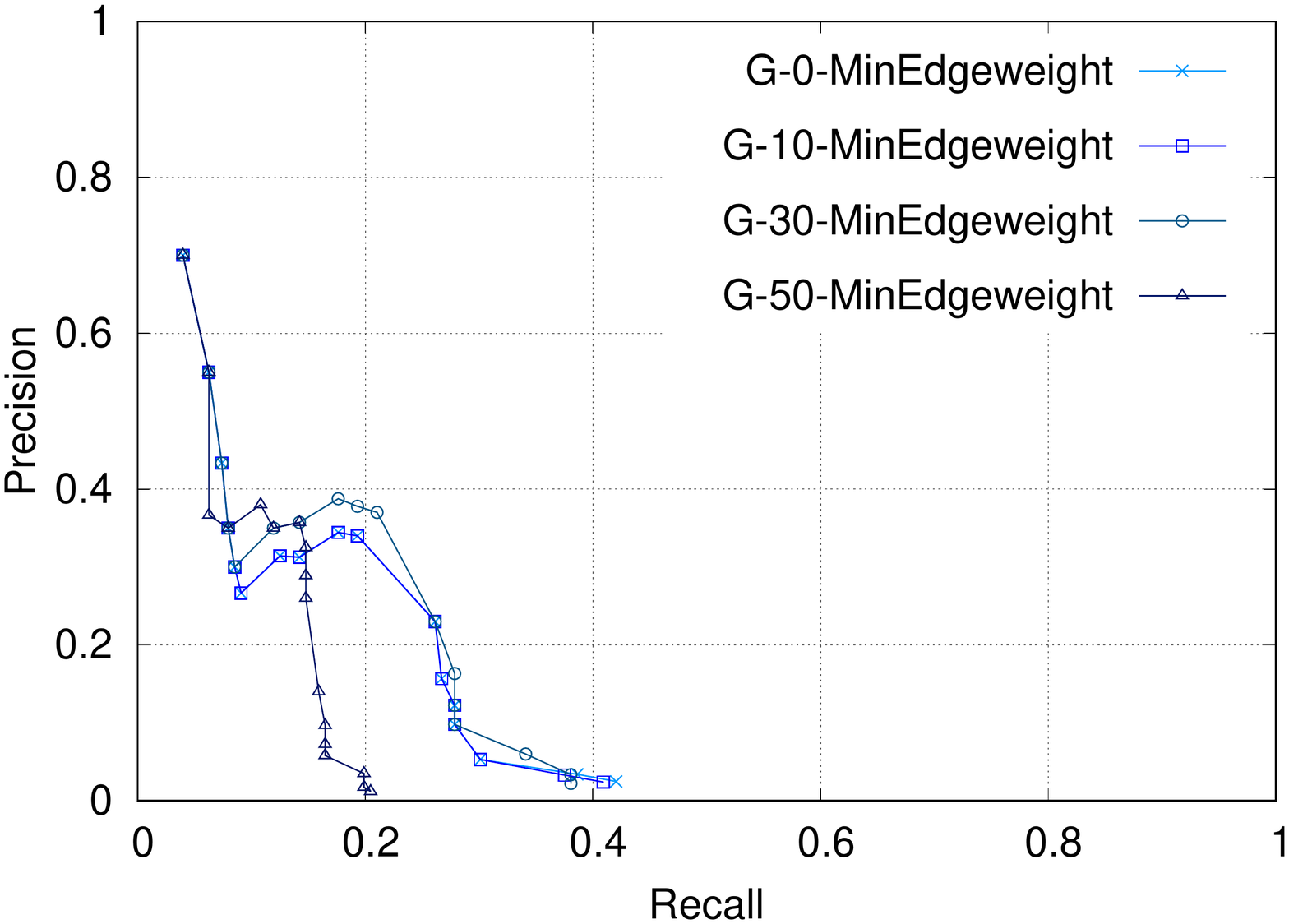}
&
\hspace{-1.6cm}
\includegraphics[scale=0.27]{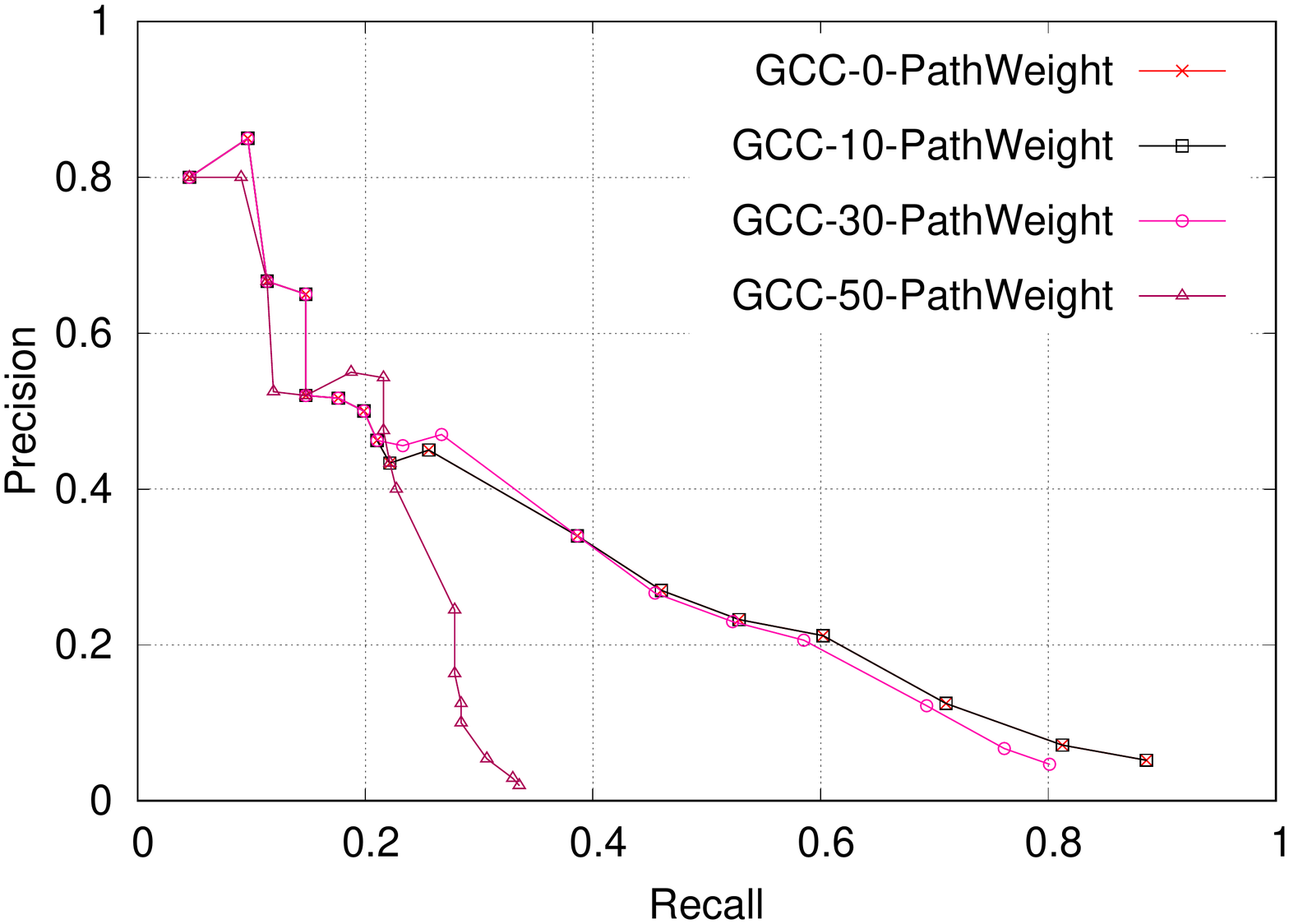}

\end{tabular}
\caption{Precision/Recall at $k$ using different values for protein complex coverage using best path metrics with Gitter method (G) and our method with cell cycle and protein complex coverage (GCC). }
\label{prec-recall-pars}
\end{figure}

We also analyze the effect of the path ranking metrics as well as the protein complex coverage ratio ($cv$) using precision and recall at $k$ for both methods.  
Figure~\ref{prec-recall-pars} shows how precision and recall at $k$ varies when increasing $k$ using Gitter (G-cv\%) with protein complex coverage and different path metrics and using our approach, Gitter with cell cycle and protein complex coverage (GCC-cv\%). Best path ranking metrics are Path Weight and Min Edge Weight for both methods.

\begin{figure}
\hspace{-5mm}
\begin{tabular}{p{0.32\textwidth}p{0.30\textwidth}p{0.31\textwidth}}
\includegraphics[scale=0.32]{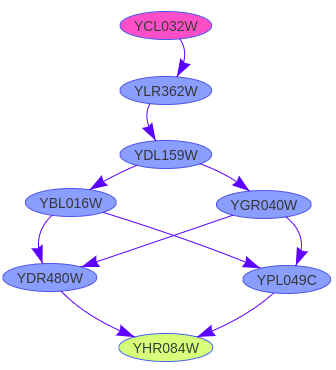}
&
\includegraphics[scale=0.32]{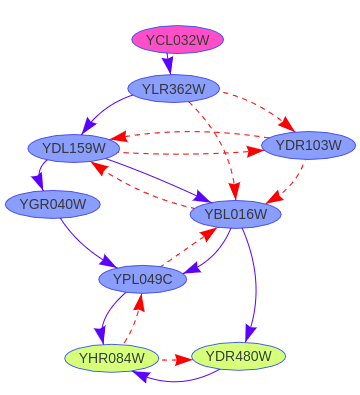}
&
\includegraphics[scale=0.29]{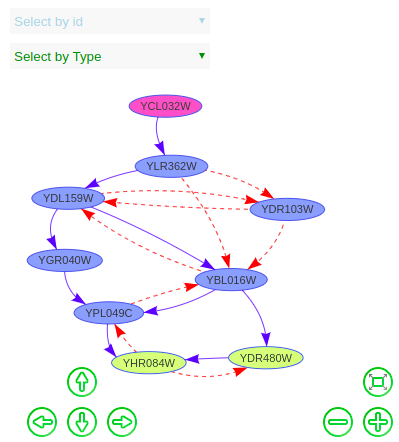}
\end{tabular}
\caption{Top20 pathways from our GCC-10\% method. Left: Pathways completely matched. Middle: Pathways partially matched, with 3 or more correct interactions. Right: visualization panel.}
\label{an}
\end{figure}

\section{Visualization}
In order to help the analysis of predicted pathways, we developed a visualization tool that processes the predicted pathways and displays them as a directed graph.
This tool was developed using visNetwork package in R, and displays the pathways in HTML format so that they can be seen through any browser. 

For example, Figure \ref{an} displays the top-$20$ predicted pathways. There are 4 completely matched predicted pathways in the top-$20$ (Figure \ref{an}-left), and there are 13 partially matched pathways in the top-$20$ (Figure \ref{an}-middle). All of them share the same source protein (YCL032W) but they have different targets (YLR362W, YDR103W). We observe that the interactions from YLR362W to YDR103W and from YDR103W to YDL159W, are the most repeated in the 13 predicted pathways from Figure \ref{an}-middle, which can be an indicator of a high level of certainty. 

As it can be seen in Figure \ref{an}-right, it can distinguish the reference from the predicted pathways,  
displaying true interactions in continue lines, and interactions that are not in the gold standard in dashed lines.
Also, users can select for highlighting individual proteins, or a group of proteins having a common attribute. 
Other allowed actions are zooming in and out, moving around proteins (vertices) and showing information about each protein.

\section{Discussion}

Analyzing Figure~\ref{an}, Ste50, the protein product of YCL032W, is an adaptor protein that mediates cell signaling between G protein-associated kinases and mitogen-activated protein kinases (MAPK)~\cite{uno}. 
In particular, it has been documented to participate in at least three different pathways controlling pheromone response/mating~\cite{dos,tres}, high osmolarity response (HOG pathway)~\cite{cuatro,cinco} and starvation/filamentation response~\cite{seis}. 
The pheromone response pathway induces mating behavior, the cell wall stress pathway produces cell wall remodeling and the starvation pathway causes filamentation. Additionally, the pheromone response pathway is indirectly related to the cell wall stress signaling pathway through Cdc42. Both these pathways and the HOG pathway can induce cell cycle arrest.
Ste50 operates as an adaptor protein that binds to Cdc42, active phosphorylated Ste20 and Ste11 in order to bring them together on the cell membrane and cause the phosphorylation of Ste11 by Ste20~\cite{tres}. In this function, Ste50 also binds to the Opy2 membrane anchor, depending on Opy2 and Ste50 phosphorylation states, which modulates the transmission of signals to the HOG or the pheromone/mating pathway~\cite{siete}.

All four MAPK pathways in yeast are interconnected in a complex way and share common elements, including Ste50~\cite{seis,ocho}. Pathways are also highly interconnected with shared elements, and heavy crosstalk is established between them. The final response triggered by the common Cdc42 (YLR229C), Ste50 (YCL032W) and Ste11 (YLR362W) proteins is determined by the exact phosphorylated sites and proteins in the cascades~\cite{siete}, and depends on the interaction of many different proteins and adaptors that form membrane-bound and cytoplasmic complexes. This complexity can be seen in the KEGG map of the yeast MAPK pathways (\url{https://www.genome.jp/kegg-bin/show_pathway?sce04011}). 

All together, these signaling events include many different protein-protein interactions that transmit signals with different intensities to pathway mediators such as Cdc42 (YLR229C), Ste11 (YLR362W), Ste7 (YDL159W), and Ste 5 (YDR103W), and pathway effectors such as Dig1 (YPL049C), Dig2 (YDR480W), Fus3 (YBL016W) and Ste12 (YHR084W) in the pheromone/mating pathway, and Kss1 (YGR040W) in the starvation/filamentation pathway.

In reference to Figure~\ref{an} as an example, our approach to discover signaling pathways completely predicted the pathways that originate in Ste50 (YCL032W) and end in one of the effectors for the pheromone/mating and starvation/filamentation pathways that were present in the reference (Figure~\ref{an}-left). This approach also allowed to discover pathways of signaling that include the Ste5 protein (YDR103W), which are not included in the reference and were therefore partially matched (Figure~\ref{an}-middle). These predicted pathways have not been reported before as signaling pathway by themselves. 
However, this is not a spurious artifact, since Ste5 has been reported to simultaneously interact as a scaffolding protein with Ste11, Ste7 and Fus3, tethering them to the membrane and forcing their correct interaction, thus increasing the signaling through the pheromone/mating pathway and preventing erroneous signaling of the upstream kinases~\cite{seis}. Also, Ste5 is necessary for Ste7 phosphorylation of Fus3 but not for phosphorylation of Kss1, which allows upstream regulation of the
 pheromone / mating pathway and directs the signaling cascade to mating rather than to filamentation~\cite{nueve,diez,once}. Therefore, the “new” pathways discovered by our approach in this case, although not in the reference, can be seen as alternative signaling pathways through which crosstalk between different pathways occurs and the final signaling response is determined. Hence, including these automatically discovered pathways in the analysis of MAPK signaling in yeast gives more detail about the information flow through pathways and are consistent with experimental results~\cite{seis,nueve,diez,once}, which further demonstrates the utility of our approach.

\section{Conclusions and future work}

We have described a method that uses biological knowledge based on cell cycle peak expression, phenotypes and protein complexes to improve prediction of signaling pathways. This approach first estimates candidate pathways using Gitter's prediction method~\cite{Gitter}. Then, it incorporates cell cycle rules by defining a graph for relations between the gene cycle stages peak expressions and phenotypes to obtain consistent pathways. Finally, it adds protein complex coverage to obtain predicted pathways.  

We evaluate our approach using path metrics to define top-$k$ rankings using different approaches. We show that our method improves Gitter's performance regarding the number of true positives in the top-100 results obtaining an improvement of 27\%. In addition, our method provides better $precision@k$, $recall@k$ and $F_1Score@k$ compared with Gitter's, PathLinker and PathLinker-RWR methods. The performance gain can be observed with the overall MAP achieved by our method, which is 0.66, compared to 0.45 obtained by Gitter's and 0.47 obtained by PathLinker-RWR.

As a future work, we plan to use gene expression clustering algorithms to incorporate a more detailed gene interaction for estimating the predicted consistent pathways as well as studying other organisms.  An idea to explore include the use of the probabilistic markov chain model in combination with gene expression clustering algorithms in order to infer protein interactions for predicting signaling pathways. Combining markov models with gene clustering have been used to infer pairwise features~\cite{mv}.

\section*{Acknowledgement} 
This research has received funding from the European Union's Horizon 2020 research and innovation programme under the Marie Skłodowska-Curie Actions H2020-MSCA-RISE-2015 BIRDS GA No. 690941

\end{document}